%% file: CHESSCUPID.tex
\documentclass[11pt,a4paper]{article}
\pdfoutput=1
\usepackage{jinstpub}

\RequirePackage{graphicx}
\RequirePackage{mathptmx}
\RequirePackage{flushend}
\RequirePackage[numbers,sort&compress]{natbib}

\usepackage{amsfonts}
\usepackage{amssymb}
\usepackage{wasysym}
\usepackage{multirow}
\usepackage{subfigure}
\usepackage{siunitx}
\usepackage{color}
\usepackage[switch]{lineno}

\RequirePackage[capitalize]{cleveref}
\crefname{figure}{fig.}{figs.}
\crefname{table}{table}{tables.}
\crefname{equation}{eqn.}{eqns.}
\crefname{section}{sec.}{secs.}

\graphicspath{{figs/}}

\include{commands}

\begin{document}

\title{Characterization of Light Production and Transport in Tellurium Dioxide Crystals}

\author[a,b]{R.~G.~Huang}
\author[a,b]{G.~Benato}
\author[a,b]{J.~Caravaca}
\author[a,b]{Yu.~G.~Kolomensky}
\author[a,b]{B.~J.~Land}
\author[a,b]{G.~D.~Orebi Gann}
\author[b]{B.~Schmidt}

\affiliation[a]{University of California, Berkeley, CA 94720-7300, USA}
\affiliation[b]{Lawrence Berkeley National Laboratory, CA 94720-8153, USA}

\emailAdd{roger\_huang@berkeley.edu}

\abstract{Simultaneous measurement of phonon and light signatures is an
effective way to reduce the backgrounds and increase the sensitivity of CUPID, 
a next-generation bolometric neutrinoless double-beta decay ($0\nu\beta\beta$) 
experiment. Light emission in tellurium dioxide (TeO$_2$) crystals, one of the
candidate materials for CUPID, is dominated by faint Cherenkov radiation, and the
high refractive index of TeO$_2$ complicates light collection. 
Positive identification of $0\nu\beta\beta$ events therefore requires 
high-sensitivity light detectors and careful optimization
of light transport. A detailed microphysical understanding of the
optical properties of TeO$_2$ crystals is essential for such
optimization. 
We present a set of quantitative measurements of light production and
transport in a cubic TeO$_2$ crystal, verified with a complete optical
model and calibrated against a UVT acrylic standard. We measure the optical surface properties of the crystal, and set
stringent limits on the amount of room-temperature 
scintillation in TeO$_2$ for $\beta$ and $\alpha$  particles of 5.3 and 8 photons / MeV, respectively, at 90\% confidence. The techniques
described here can be used to optimize and verify the particle
identification capabilities of CUPID.}
  


\arxivnumber{1907.10856}

\date{Received: date / Accepted: date}

\maketitle

\input{Introduction}

\input{Chess_description}

\input{Analysis}

\input{Results}

\section{Conclusions}
\label{sec:Conclusion}

We have presented a comprehensive set of measurements of  light
production and optical transport properties of a TeO$_2$ crystal, verified by detailed
simulations and measurements in a well-understood medium (UVT
acrylic). This establishes techniques for quantitative analysis and 
optimization of light collection in bolometric crystals, which can
be done in a conventional room temperature experimental setup. We
have demonstrated that the use of $\beta$ sources, together with a tagged cosmic muon
dataset, allows us to break degeneracies between  light
production and transport mechanisms, and we have verified that TeO$_2$ has
negligible scintillation for both $\beta$ and $\alpha$ excitation at room temperature.

Future extensions of this work will include quantitative analysis of
crystal surface treatment options, such as  polish and  coating
with reflective or refractive media, which can be used to optimize design
of the CUPID detector modules. As CHESS is designed to be able to effectively
separate Cherenkov and scintillation signals, we also plan to
measure the optical properties of scintillating crystals such as
Li$_2$MoO$_4$, which has been chosen as the baseline for the CUPID
experiment~\cite{CUPIDInterestGroup:2019inu}. 

\section{Acknowledgments}

This material is based upon work supported by the US Department of
Energy (DOE) Office of Science under Contract No. DE-AC02-05CH11231,
and by the DOE Office of Science, Office of Nuclear Physics under
Award Nos. DE-FG02-00ER41138 and DE-SC0018987.
The CHESS apparatus was constructed using funds from the Laboratory Directed Research and Development Program of Lawrence Berkeley National Laboratory under U.S. Department of Energy Contract No. DE-AC02-05CH11231.
This research used the Savio computational cluster resource provided by the Berkeley Research Computing program at the University of California, Berkeley (supported by the UC Berkeley Chancellor, Vice Chancellor for Research, and Chief Information Officer).

We thank the CUORE collaboration for providing the TeO$_2$ crystal.
We would like to thank Joe Wallig for his valuable contributions, and members of the Berkeley Weak Interaction and Underground Physics groups
for stimulating discussions.

\bibliographystyle{JHEP}
\bibliography{Bibliography}

\end{document}

%% file: commands.tex
\newcommand{\onbb}{$0\nu\beta\beta$}

\newcommand{\cupido}{CUPID-0}
\newcommand{\cupidmo}{CUPID-Mo}

\newcommand{\Mo}{$^{100}$Mo}

\newcommand{\Se}{$^{82}$Se}
\newcommand{\Cd}{$^{116}$Cd}

\newcommand{\teo}{TeO$_2$ }

%% file: Introduction.tex
\section{\label{sec:intro} Introduction}

The study of neutrinoless double beta decay (\onbb) with tellurium dioxide (TeO$_2$) crystals in the CUORE experiment is one of
the most sensitive low-energy searches for lepton number violation today \cite{Alduino:2017ehq}. 
A positive signal would prove the existence of massive Majorana neutrinos and imply lepton number violation by two units~\cite{Tanabashi:2018oca, Elliott:2002xe, AVIGNONE2008481}. 
The hypothetical decay produces a monochromatic peak in the summed kinetic energy
sum spectrum of the two emitted electrons. CUORE uses a bolometric approach to look for this signature peak at the \onbb\ transition energy, directly measuring the heat energy deposited by events in the \teo crystals.

The main limiting factor in CUORE's \onbb\ sensitivity is currently its backgrounds, primarily from degraded $\alpha$ events \cite{CUOREBackground2017}. These events deposit part of their energy in a passive part of the detector support structure and can hence produce a background that extends into the region of interest around 2527.5 keV \cite{Te130QValue}. The ability to perform active background rejection against these events is 
considered a fundamental requirement for CUPID, the next generation
upgrade to CUORE \cite{CUPIDWhitePaper, CUPIDInterestGroup:2019inu}.  

One of the most promising methods of active background discrimination for CUPID is achieved by employing a thin secondary Ge or Si wafer based bolometer, 
which is able to detect the Cherenkov and/or scintillation light from energy deposits in the crystals and distinguish between events through different light yields and pulse shapes for different particle types. This principle has already been demonstrated for several $\beta\beta$-decay candidates such as \Se, \Mo, and \Cd, most recently in \cupido~\cite{Azzolini:2018dyb} at Laboratori Nazionali del Gran Sasso (LNGS) and in \cupidmo~\cite{CUPIDMo2019} at Laboratoire Souterrain de Modane (LSM). For the target crystals under consideration for use in CUPID, the reported light yields vary by two orders of magnitude from 40-60 eV/MeV in \teo to 6.4 keV/MeV in ZnSe \cite{Barucci:2019ghi, Pattavina2016, Beeman:2013vda}. Depending on which type of crystal is chosen for CUPID, detector design may have to be optimized for either low light yields or very high light yields.

Experimental tests of different detector designs can be both costly and time consuming, 
as real size detector tests often need to be carried out in dilution refrigerator setups 
occupying valuable laboratory space at deep underground sites like LNGS or LSM. 
In this work we present a complementary approach in which we combine measurements in a room temperature setup called CHESS \cite{Caravaca2017a} with a detailed microphysical model of the apparatus in order to extract certain properties of the \teo crystal, including both the electron and alpha scintillation yields. The resulting validated Monte Carlo model, built in a framework called RAT-PAC \cite{Ratpac}, can  be used to inform and optimize next-generation detector design. 

Light emission from TeO$_2$ is dominated by Cherenkov light \cite{Casali2017} 
with a minor luminescence component reported in some cryogenic measurements \cite{Berge2018, Coron2004}. 
Particle discrimination in \teo thus relies on the detection of the small amounts of Cherenkov light from $\beta,\gamma$ events versus no light from $\alpha$ events 
that have energies below the Cherenkov light production threshold. 
Given the small amount of light produced, it is important to understand the optical properties of the crystal and be able to maximize the light collection efficiency. Previous efforts to better understand the optical properties of TeO$_2$ \cite{Casali2013, Bellini2014, Casali2017} employed a package developed for use in the CMS experiment around 2009/2010 called Litrani \cite{Litrani} and found that modest improvements to the light yield could be achieved with a combination of rougher surfaces of the TeO$_2$ crystals and better matching of the light detector geometry to the cubic crystals.

The simulation package we employ has previously been used in \cite{Caravaca2017, Caravaca2017a} 
to model and demonstrate the possibility of separating Cherenkov and scintillation light production 
on an event-by-event basis in liquid scintillators.
It is based on modern software packages and is readily adaptable to different target crystals. 
In this work, we perform a first quantitative analysis of any room temperature scintillation-like light in TeO$_2$ and demonstrate the validity of this simulation framework against data taken in the CHESS setup

We start by introducing the CHESS setup in Sec. \ref{sec:setup}. The measurements were taken in two distinct configurations, described in Secs. \ref{sec:setup_muon} and \ref{sec:setup_source}, which we use to collect cosmic-muon, $\beta$, and $\alpha$ data.
We model the entire setup in GEANT4 (Sec. \ref{sec:setup_MC}) and perform a combined likelihood analysis in which we determine the best model parameters in terms of the crystal roughness and the number of surplus scintillation photons per MeV of energy deposit.  
This analysis is detailed in Sec. \ref{sec:Analysis} and validated on a mock-up target of ultra violet transparent (UVT) acrylic. 
The results with our limits on $\beta$ and $\alpha$ scintillation in \teo are discussed in Sec. \ref{sec:Results}, and concluding remarks and future directions are given in the conclusion (Sec. \ref{sec:Conclusion}).

%% file: Chess_description.tex
\section{The CHESS apparatus}
\label{sec:setup}

We use an existing apparatus called CHESS for these measurements. CHESS is a bench-top optical detector that features a cross-shaped array of 12 1-inch cubic PMTs (H11934-200 from Hamamatsu) facing a target that can be exposed to cosmic muons, a beta source ($^{90}$Sr), or an alpha source ($^{241}$Am). In our case, the target is a solid 5-cm sized \teo cube that will produce Cherenkov radiation, and potentially scintillation light, in response to the passage of charged particles.  In between the PMT array and the target we place an acrylic block serving as an optical propagation medium to reduce internal reflections in the target. The block is optically coupled to the PMTs and target using Eljen optical grease (EJ-550). This setup is surrounded by 4 scintillator panels used as vetos, providing $4\pi$ coverage from through-going cosmic rays. We use two different configurations for muon and radioactive-source measurements. These configurations are described in the next sections, and further details can be found in \cite{Caravaca2017a}.

\begin{figure}
	\centering
	\includegraphics[width=0.9\columnwidth]{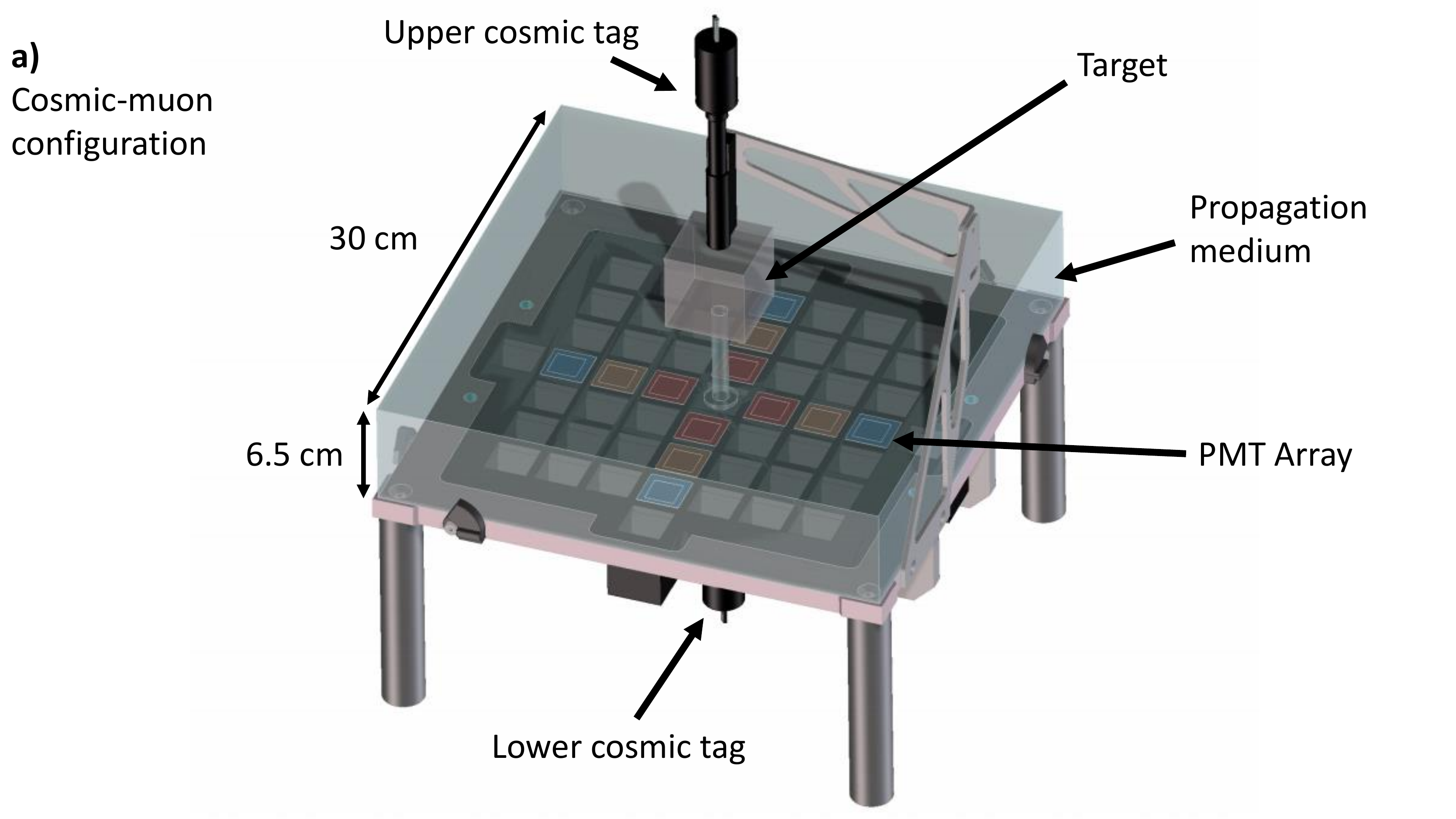}
	\includegraphics[width=0.9\columnwidth]{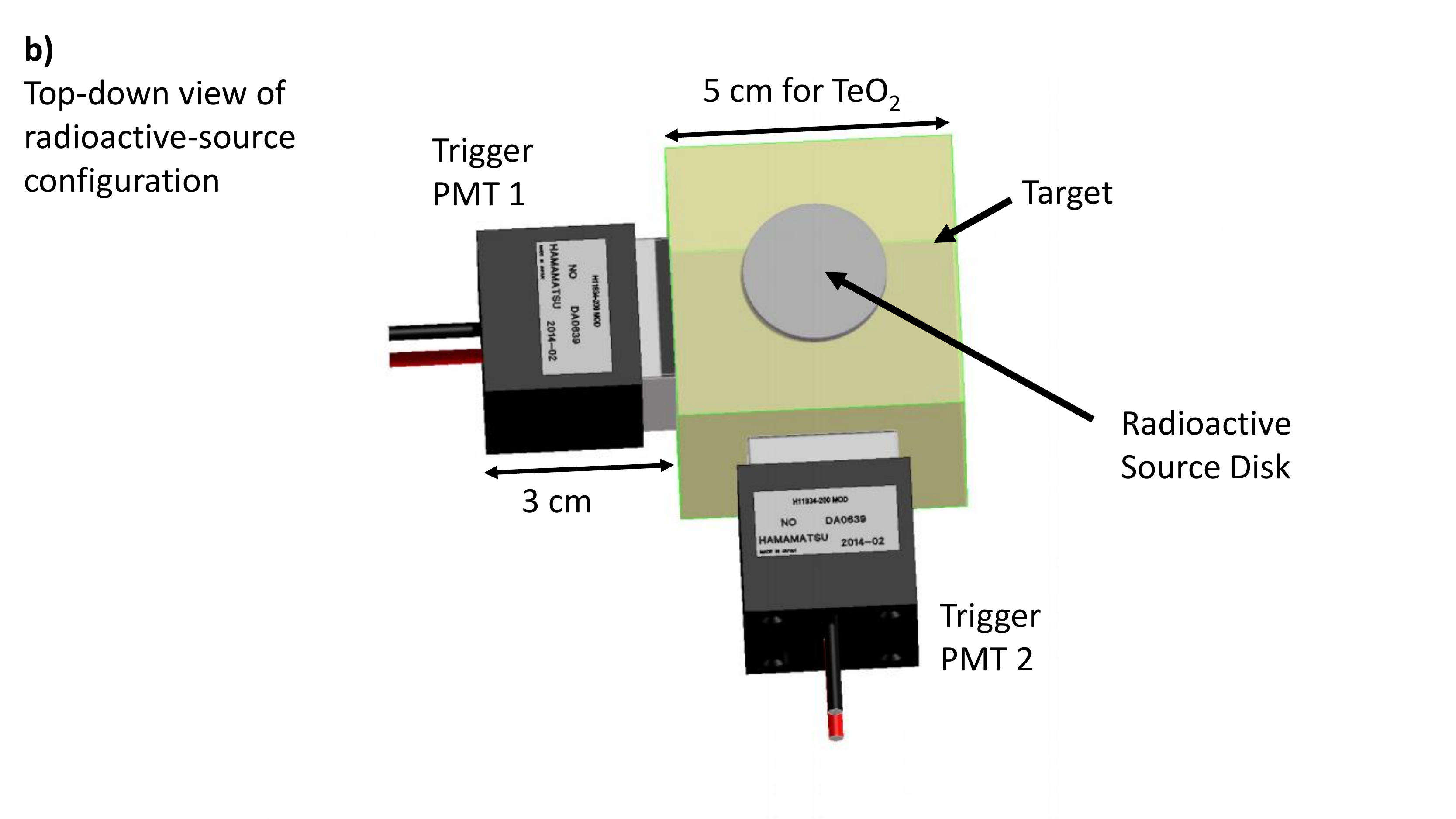}
	\caption{a) The CHESS apparatus in the cosmic-muon configuration (top), and b) in the radioactive-source configuration (bottom). The array shown in (a) contains 53 potential slots for PMTs in total, but only the 12 colored slots contain PMTs for this work. These 12 PMTs are color-coded by their radial distance from the center of the setup, with the inner ring in red, the middle ring in orange, and the outer ring in blue. The main difference between (a) and (b) is the substitution of the muon trigger system by the trigger PMTs attached to the target. In the radioactive-source configuration, the 12 PMTs beneath the target are not used.}
	\label{fig:chess}
\end{figure}

\subsection{Cosmic-muon configuration}
\label{sec:setup_muon}

The configuration used for cosmic-muon data collection is shown in Fig. \ref{fig:chess}a, in which we trigger on downward-going muons and measure in the set of 12 PMTs beneath the target the spatial distribution of any light produced. This configuration is particularly sensitive to the surface properties of the target as any roughness will affect both refraction and reflection at the surface, modifying the spatial distribution of light incident on the PMT array. We detect vertical cosmic muons with a triple coincidence technique, requiring a signal in two cylindrical muon scintillator tags located above and below the target and PMT array, as well as in a flat scintillator panel immediately underneath the setup. Shower events with more than a single muon are rejected by vetoing events with a signal in any of the other scintillator panels. The acrylic block includes a hollow of the same radius as the muon tags in order to prevent muons from producing Cherenkov radiation in the acrylic itself. We record PMT data for all events that trigger the bottom muon tag and require the triple coincidence during offline analysis.

\subsection{Radioactive-source configuration}
\label{sec:setup_source}

The configuration used for radioactive-source data collection is shown in Fig. \ref{fig:chess}b. We deploy either a $^{90}$Sr or $^{241}$Am source directly on top of the target and optically couple two cubic PMTs of the same type as those in the PMT array to two faces of the target. We configure the hardware to trigger on one of the PMTs, referred to as trigger PMT 1, and require an offline double coincidence with the other PMT, referred to as trigger PMT 2. The hardware trigger threshold is set to be about 2 photoelectrons (PEs), selected to be well clear of the electronic noise peak. In the offline selection, we require at least 3 PEs in trigger PMT 1 to allow us to ignore any trigger threshold effects and require at least 1 PE in trigger PMT 2 for the double coincidence. This double coincidence scheme helps cut spurious events caused by dark noise on trigger PMT 1. Cosmic rays are vetoed by requiring that no signal is observed in any of the scintillator panels.

\subsection{Monte Carlo model}
\label{sec:setup_MC}

We perform a GEANT4-based Monte Carlo simulation using the RAT-PAC toolkit \cite{Ratpac}, which is able to do single-photon tracking while taking into account effects from geometry, optical properties of different media, and the photon detector and DAQ system properties. All of these properties and the relevant physics processes are simulated as described in \cite{Caravaca2017,Caravaca2017a}. Simulation of the \teo crystal requires the material composition, which is extracted from \cite{ChemicalWeights}, as well as its refractive index and optical absorption lengths, both extracted from \cite{Casali2017}. \teo is also known to be a birefringent material, which we do not have the ability to fully simulate in our software framework. We tested the potential effects of birefringence by varying the index of refraction in \teo from the ordinary to extraordinary values in our simulations; the effects were found to be negligible in all of our setups and were ignored in the rest of the analysis.

To account for the possibility of scintillation in TeO$_2$, we use the GLG4Scint model implemented in RAT-PAC. We introduce a light yield parameter which controls the number of photons produced per MeV of energy deposited in the crystal. The spectrum and time profile of the emitted scintillation photons in our model are arbitrary, given that there is no previous measurement. These do not affect our current analysis since we are interested in determining the presence or absence of a scintillation signal, rather than its characteristics, and so our measurements are not sensitive to the details of the emission process. We set the emission spectrum to be in the sensitive region of our PMTs and we model the time profile as a simple exponential decay with a decay constant of 3.8 ns.

The cubic \teo crystal used in this experiment has 2 glossy faces and 4 matte faces, which in general cannot be treated as perfectly smooth optical surfaces. We account for this effect using the GLISUR model \cite{Geant4:Manual} and a polish parameter that can be varied from 0 to 1, with 1 corresponding to normal Snell's law refraction/reflection and 0 corresponding to a completely diffuse surface. We assign one polish value to the glossy faces and one value to the matte faces of the TeO$_2$ crystal. The polish and light yield are treated as free parameters in our model and are obtained by fits to the data.

\subsection{Calibration}
\label{sec:setup_Calibration}

We calibrate the setup with a UVT acrylic target, which is well understood and can be precisely modeled. Individual PMT gains are measured by deploying a $^{90}$Sr source on top of the acrylic target, which provides a charge distribution with a distinct single photo-electron (SPE) peak for each PMT. We fit the resulting spectrum in the same way as in \cite{Caravaca2017a}, using a Gaussian with mean $\mu_e$ and standard deviation $\sigma_e$ for the noise peak and including the multi-PE event distributions up to 3 PEs, with each $n$ PE peak being a Gaussian of mean $\mu_e + n*\mu_{SPE}$ and standard deviation $\sqrt{\sigma_e^2 + n*\sigma_{SPE}^2}$. The SPE Gaussian parameters extracted from this fit then serve as our PMT charge model. To check for deviations in the PMT gain caused by coupling to the trigger hardware, we perform the SPE calibration for trigger PMTs 1 and 2 three times using three slightly different trigger methods in the radioactive-source configuration: two calibrations are done by triggering on trigger PMT 1 and 2, respectively, and the last one is done by triggering on a third PMT that was temporarily coupled to the acrylic target for this purpose. Calibrations were consistent across all three datasets within uncertainties.

%% file: Analysis.tex
\section{Analysis methods}
\label{sec:Analysis}
In our analysis we consider both the cosmic muon data collected using the cosmic-muon configuration and the $\beta$ and $\alpha$ data collected in the radioactive-source configuration. Background data is collected for the radioactive-source configuration by taking data in the absence of a source with the same triggering system. The entire analysis chain is tested with a UVT acrylic mock-up target of similar dimensions (6x6x6 cm$^3$) to the TeO$_2$ crystal. Since the optics of the acrylic sample are well understood, this provides a validation of the simulation framework and analysis methods, as well as an estimate of certain systematic uncertainties. The analysis is then repeated for the \teo crystal.

\subsection{Electron scintillation yield}
\label{sec:BetaMuonAnalysis}

We use the muon and electron data to measure the quantity of scintillation light caused by any electron-like energy deposition, which we parametrize in the Monte Carlo model as $\ell$ scintillation photons / MeV. The amount of light we can detect coming from the target also depends on the polish of the surfaces, which we parametrize as two polish parameters, $p_1$ and $p_2$, corresponding to the glossy and matte faces of the \teo crystal. For the UVT acrylic analysis it is assumed that the cube is isotropic, and we instead use just one polish parameter, $p$.

In order to disentangle the effects of scintillation yield and surface polish on the amount of light we expect to observe, the analyses for the cosmic-muon configuration and the radioactive-source configuration are performed separately and then combined. The two datasets have different sensitivities to these two sets of parameters, allowing us to break the correlation present in a single dataset. We use our Monte Carlo model to simulate the expected results for both configurations while scanning over the parameters $(p_1, p_2, \ell)$ and use these to calculate the likelihood values $P$($\mu \text{ Data} \vert p_1, p_2, \ell$) and $P$($\beta \text{ Data} \vert p_1, p_2, \ell$). The sum of the two negative log likelihoods is minimized to determine the best-fit polish and scintillation yield parameters.

For the cosmic muon data, we consider on an event-by-event basis the ratios between the numbers of photoelectrons observed in each of the three radial groupings of PMTs in the array (see Fig. \ref{fig:chess}a), for a total of two independent ratios. For these ratios, we perform a Kolmogorov-Smirnov (KS) test between the Monte Carlo predicted distribution and the observed distribution. We use a toy Monte Carlo to generate the expected distribution of KS test results given our available statistics and use this distribution to convert the KS test results into likelihood values.

The primary source of systematic uncertainty for the cosmic muon data is potentially unequal light collection efficiencies in the 3 radial PMT groupings due to geometrical effects. We account for this by re-running our cosmic-muon Monte Carlo for the acrylic target while varying the light collection efficiencies for each radial grouping of PMTs and checking the effects on the fit quality for the acrylic data. We reject any configurations that worsen our best fit quality by more than 1.35 log likelihood units (the 90\% confidence level) or that exclude the expected true value of 0 scintillation in acrylic at more than the 90\% level. After these cuts, there remains a 15\% uncertainty in light collection efficiency, which is treated as a systematic uncertainty for the \teo cosmic muon data analysis. We assume a uniform probability distribution over these light collection efficiencies and repeat the TeO$_2$ analysis for each of them, obtaining the contribution from this uncertainty by averaging the likelihoods from the repeated analyses.

For the $\beta$ data, we consider the total number of observed events that pass the cuts described in Sec. \ref{sec:setup_source} and compare it against the Monte Carlo predicted number of events, yielding a likelihood according to Poisson statistics. This comparison is sensitive to a number of systematic uncertainties, whose magnitudes are summarized in Table \ref{table:SrSystematics}. Constraints on the $^{90}$Sr source activity come from an independent measurement performed with a CZT detector, in which we looked at the observed energy spectrum and total counts. Trigger efficiency effects come from a 29-ms deadtime built into our DAQ system after each trigger, which produces some inefficiency dependent on the event rate. This is calculated separately for each dataset by looking at the time between triggers before we apply any offline cuts and extrapolating to determine how many events are missed during the deadtimes. These efficiencies come out to be $(6.4 \pm 0.1)\%$ for acrylic $\beta$ data, $(10.6 \pm 0.1)\%$ for TeO$_2$ $\beta$ data, and $(24.5 \pm 0.1)\%$ for background data. We account for the effects of these systematics by repeating the analysis 1000 times while sampling these parameters from Gaussian distributions; the likelihoods for each possible light yield are then given by the average of the results obtained from these repeated analyses.

\begin{table}
\centering
\caption{Systematic uncertainties that affect the $\beta$ and $\alpha$ analyses. Constraints on the $^{241}$Am source activity come from the manufacturer's specifications. The other systematics are described in Sec. \ref{sec:BetaMuonAnalysis}.}
\begin{tabular}{l|l}
Source of Systematic & Fractional Uncertainty                                                               \\ \hline
SPE calibration (trigger PMT 1)                                                   & 1.7\%                                                                              \\
SPE calibration (trigger PMT 2)                                                    & 0.6\%                                                                              \\
$^{90}$Sr source activity                                             & 0.8\%                                                                              \\
$^{241}$Am source activity                                              & 3.1\%                                                                              \\ \hline
Trigger efficiency                                                         & \begin{tabular}[c]{@{}l@{}}1.6\% for UVT acrylic\\ 0.9\% for TeO$_2$ \\ 0.4\% for Background\end{tabular}
	\end{tabular}
\label{table:SrSystematics}
\end{table}

To obtain the final result, the negative log likelihoods from the cosmic muon and $\beta$ analyses are summed and profiled over the polish parameters to give $P$($\mu, \beta \text { Data} \vert \ell$), a function only of the scintillation light yield $\ell$ added to the Monte Carlo. For each light yield $\ell$, the effects of the cosmic muon analysis and $\beta$ analysis systematic uncertainties are independently combined to give the total likelihoods, which we then use to set a limit on the amount of electron scintillation compatible with our data.

\subsection{Alpha scintillation yield}

We analyze the $\alpha$ data independently of the $\beta$ data, as the scintillation yields for $\beta$ and $\alpha$ particles are in general not the same for any particular material, and we expect $\alpha$ light yields to be lower due to the absence of Cherenkov light. In our simulations for the $\alpha$ setup, we fix the polish parameters to the values obtained from the electron and muon analysis best fit and only scan the scintillation light yield $\ell$. We subject the $\alpha$ data to the same cuts on the two trigger PMTs as in the $\beta$ analysis and compare the background-subtracted rate of events passing these cuts against the MC-predicted rates for various scintillation yields to set a limit on the amount of $\alpha$ scintillation in \teo.

Due to the low event rate and long length of the $\alpha$ and background datasets, we account for the possibility of fluctuations in the event rates over time by splitting the datasets into $n$ time chunks. We then introduce onto the total average rate a systematic uncertainty of 1 /$\sqrt{n}$ times the standard deviation of the rate over time. Quantified in this manner, the variations in the rates over time yield a less than 1\% effect, though it is still greater than the expected variations from purely statistical effects.

%% file: Results.tex
\section{Results}
\label{sec:Results}
\subsection{Electron scintillation yield}

\begin{figure}
  \centering
  \includegraphics[width=0.7\columnwidth]{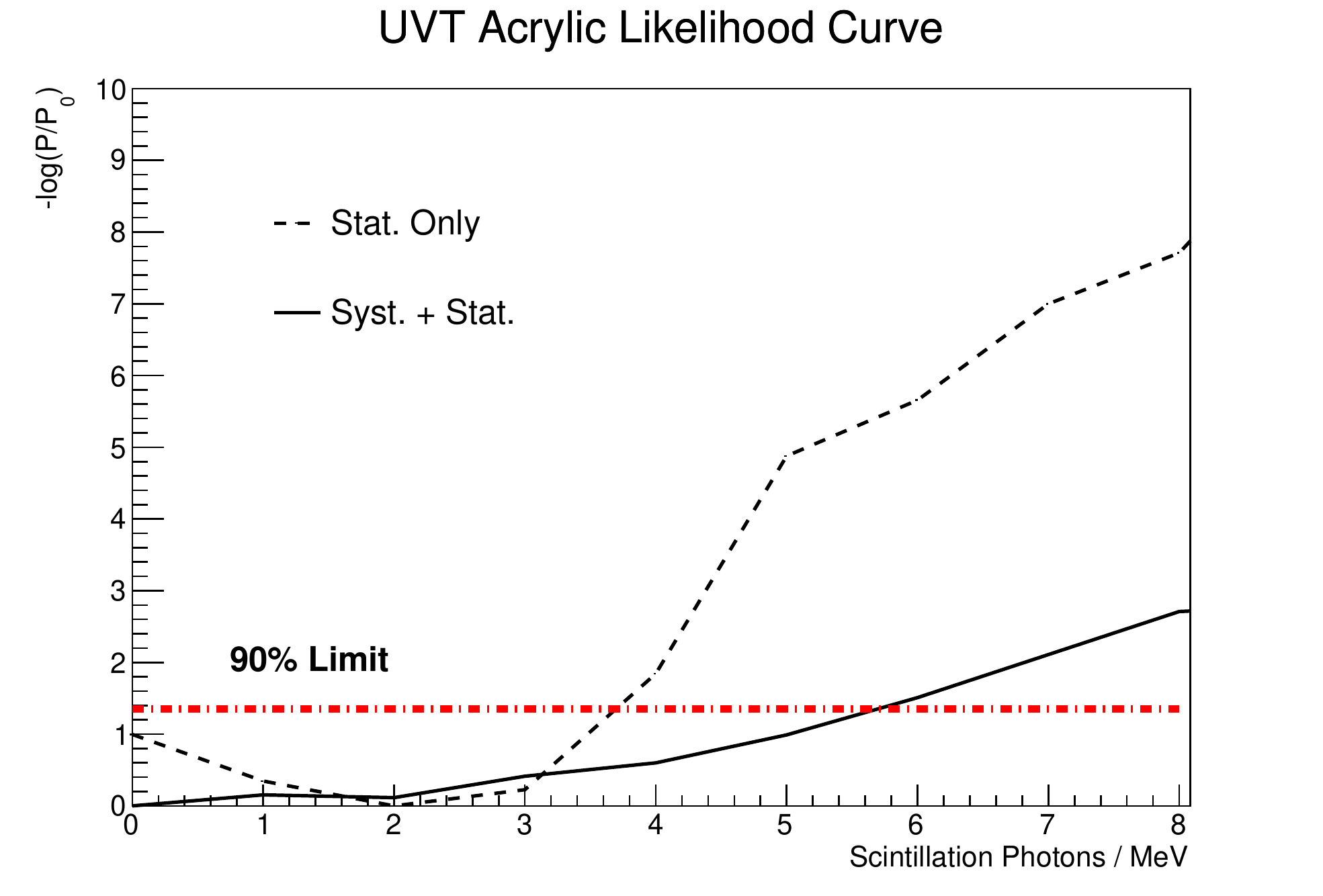}
  \caption{Negative log likelihood curve for the scintillation yield of UVT acrylic plotted both with and without systematic uncertainties, normalized to have a best fit of 0 log likelihood. Our result is consistent at the 90\% confidence level with the expected true value of 0 scintillation.}
  \label{fig:UVT_NLL}
\end{figure}

\begin{figure}
  \centering
  \includegraphics[width=0.65\columnwidth,page=1]{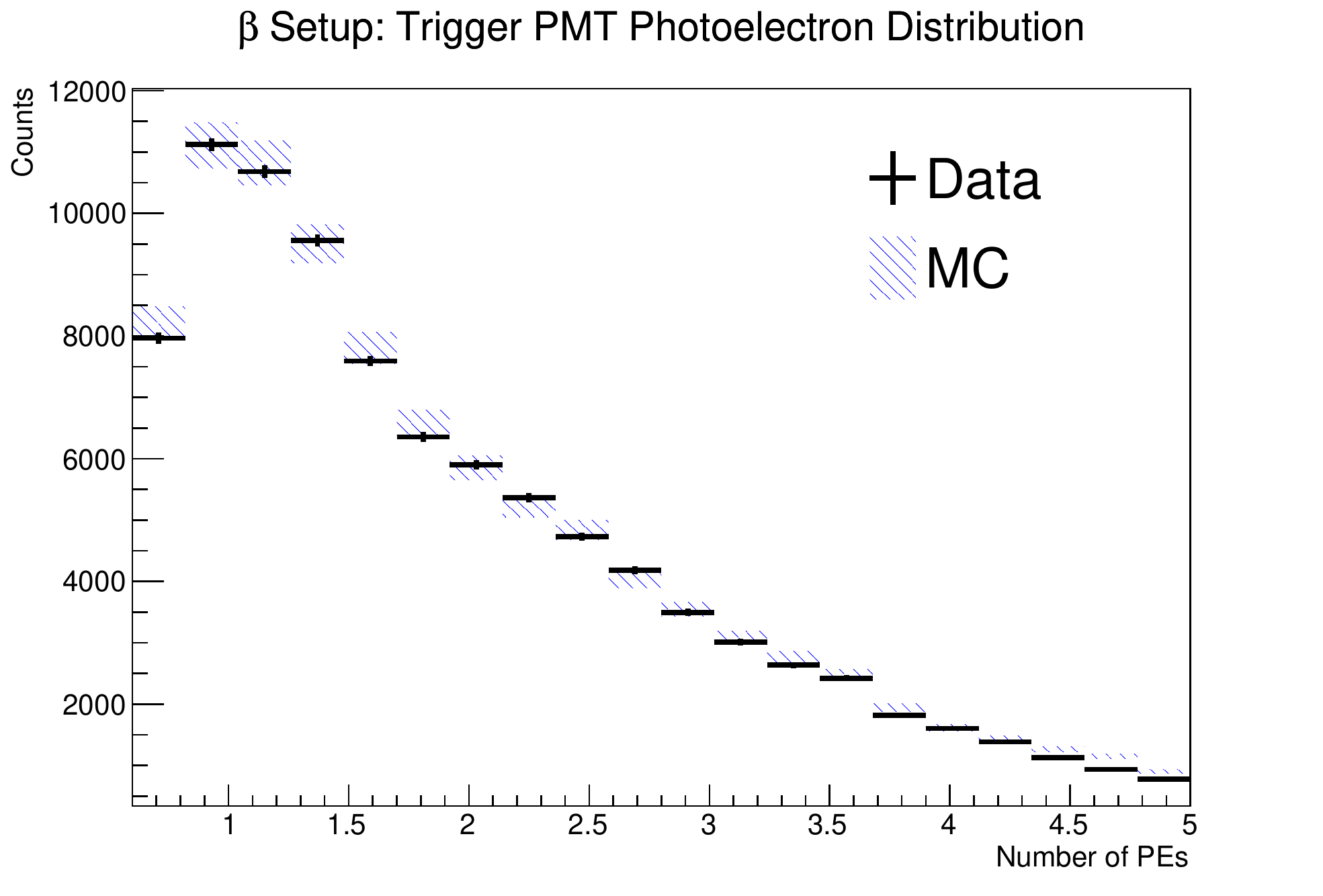} \\
  \includegraphics[width=0.45\columnwidth,page=2]{figs/UVTBestFits.pdf}
  \includegraphics[width=0.45\columnwidth,page=3]{figs/UVTBestFits.pdf}
  \caption{Distributions for the nominal best fit (statistical error only) for the UVT acrylic $\mu$/$\beta$ analysis. This corresponds to a polish of 0.86 and scintillation yield of 2 photons / MeV, which can be seen to minimize the negative log likelihood in Fig. \ref{fig:UVT_Contour}. Shaded Monte Carlo regions correspond to $1\sigma$ systematic uncertainties. There are no systematics on the cosmic muon comparison here because the acrylic data is used to determine the cosmic-muon configuration systematics. Top: PE distribution in trigger PMT 2, used for $\beta$ event counting. Bottom left: Ratio of number of PEs in the middle ring to the number in the inner ring for cosmic muons. Bottom right: Ratio of number of PEs in the outer ring to the number in the middle ring for cosmic muons.}
  \label{fig:UVT_BestFits}
\end{figure}

\begin{figure*}
  \centering
  \includegraphics[width=0.32\textwidth]{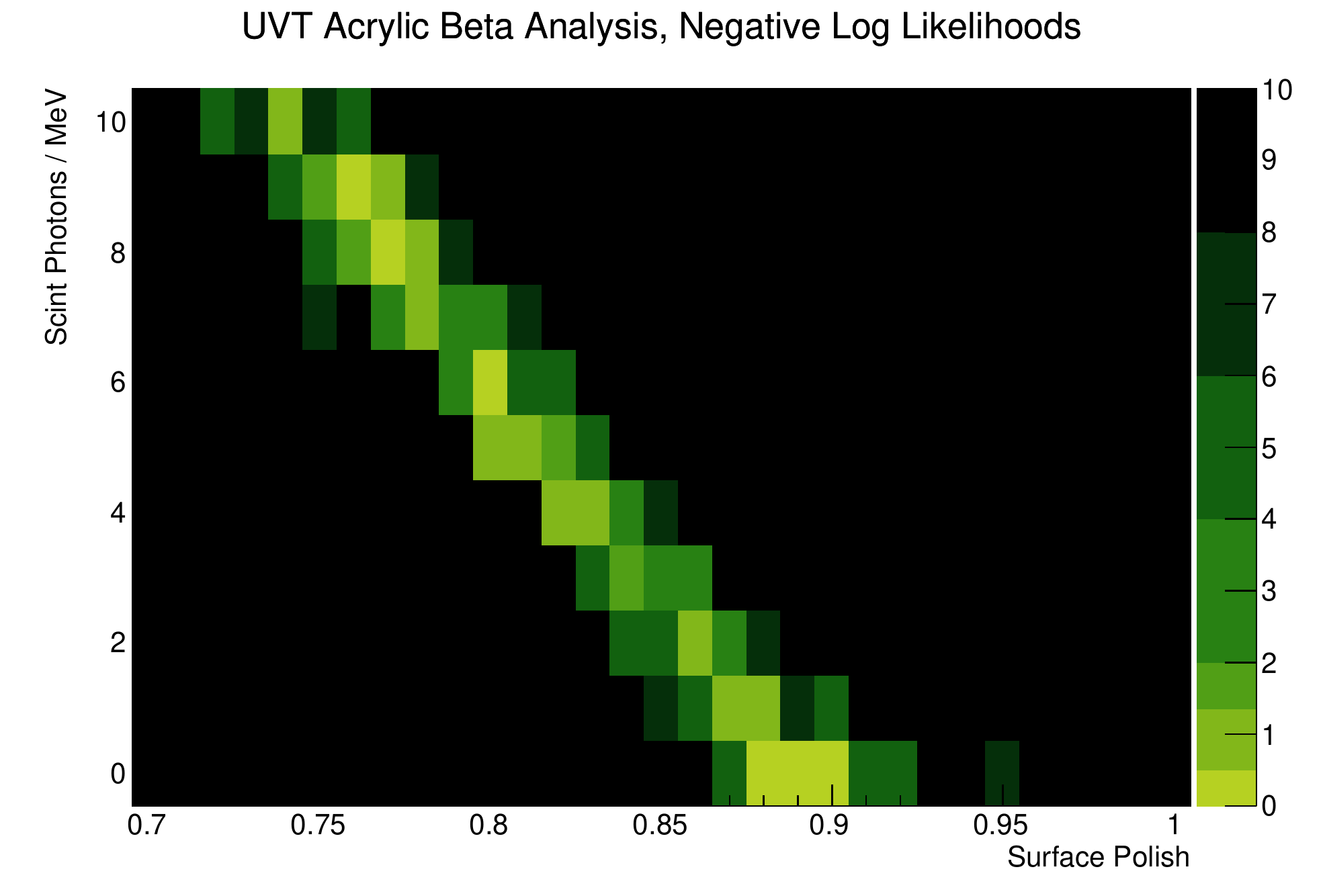}
  \includegraphics[width=0.32\textwidth]{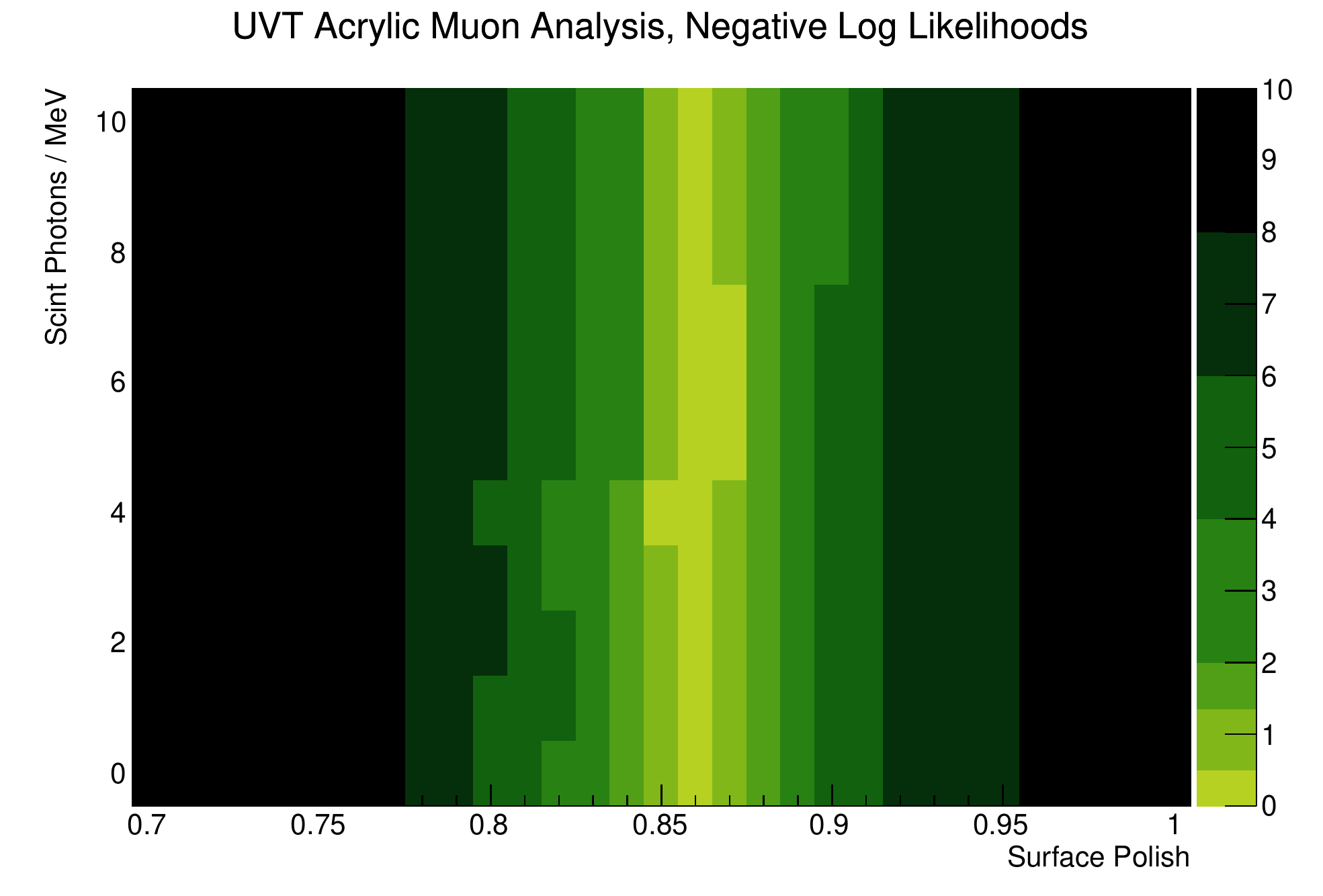}
  \includegraphics[width=0.32\textwidth]{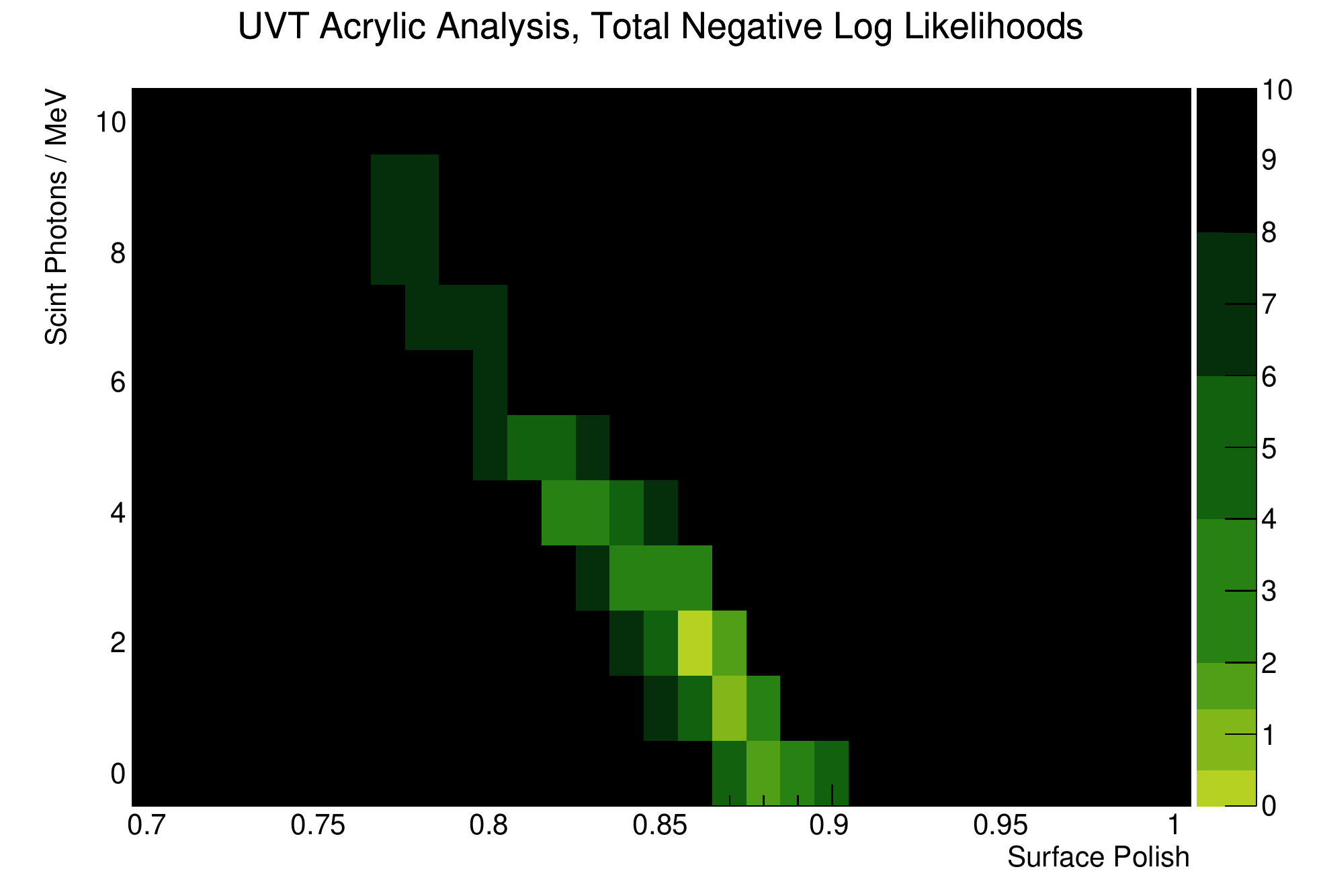}
  \caption{Left: negative log likelihood of the $\beta$ data for various polish and scintillation settings in UVT acrylic, where it can be seen that there is some degeneracy in the effects of decreased polish and increased scintillation yield. Middle: negative log likelihood of the muon data for UVT acrylic, showing high sensitivity to the polish parameters but little sensitivity to small additions of scintillation light. Right: the total negative log likelihoods obtained by combining the $\beta$ and muon analyses. Each plot is normalized to have a minimum negative log likelihood of 0 and does not include systematic uncertainties.}
  \label{fig:UVT_Contour}
\end{figure*}

The resulting delta log likelihood curve for the scintillation yield of UVT acrylic is shown in Fig. \ref{fig:UVT_NLL}, corresponding to a 90\% CL upper limit of 5.7 scintillation photons / MeV and demonstrating consistency with 0 scintillation. The data overlaid with the Monte Carlo-predicted distributions for the nominal best fit polish and scintillation yields are shown in Fig. \ref{fig:UVT_BestFits}. We use the fact that no scintillation is expected from acrylic to serve as a check on the validity of our Monte Carlo model and analysis methods and to put limits on our systematic uncertainties as described in Sec. \ref{sec:BetaMuonAnalysis}. The contributions of the $\beta$ and muon components of the analysis are shown in Fig. \ref{fig:UVT_Contour}, where it can be seen that the cosmic-muon configuration in CHESS provides good sensitivity to the surface polish parameter but has relatively poor sensitivity to small excesses of scintillation light, whereas the $\beta$ data in the radioactive-source configuration provides finer sensitivity to excess scintillation light but cannot completely distinguish between the effects of an added scintillation light yield and adjustments to the surface polish of the target. These complementary sensitivities allow us to make a precise determination of both the scintillation and polish parameter by combining the muon and $\beta$ data. 

\begin{figure*}
  \centering
  \includegraphics[width=0.32\textwidth]{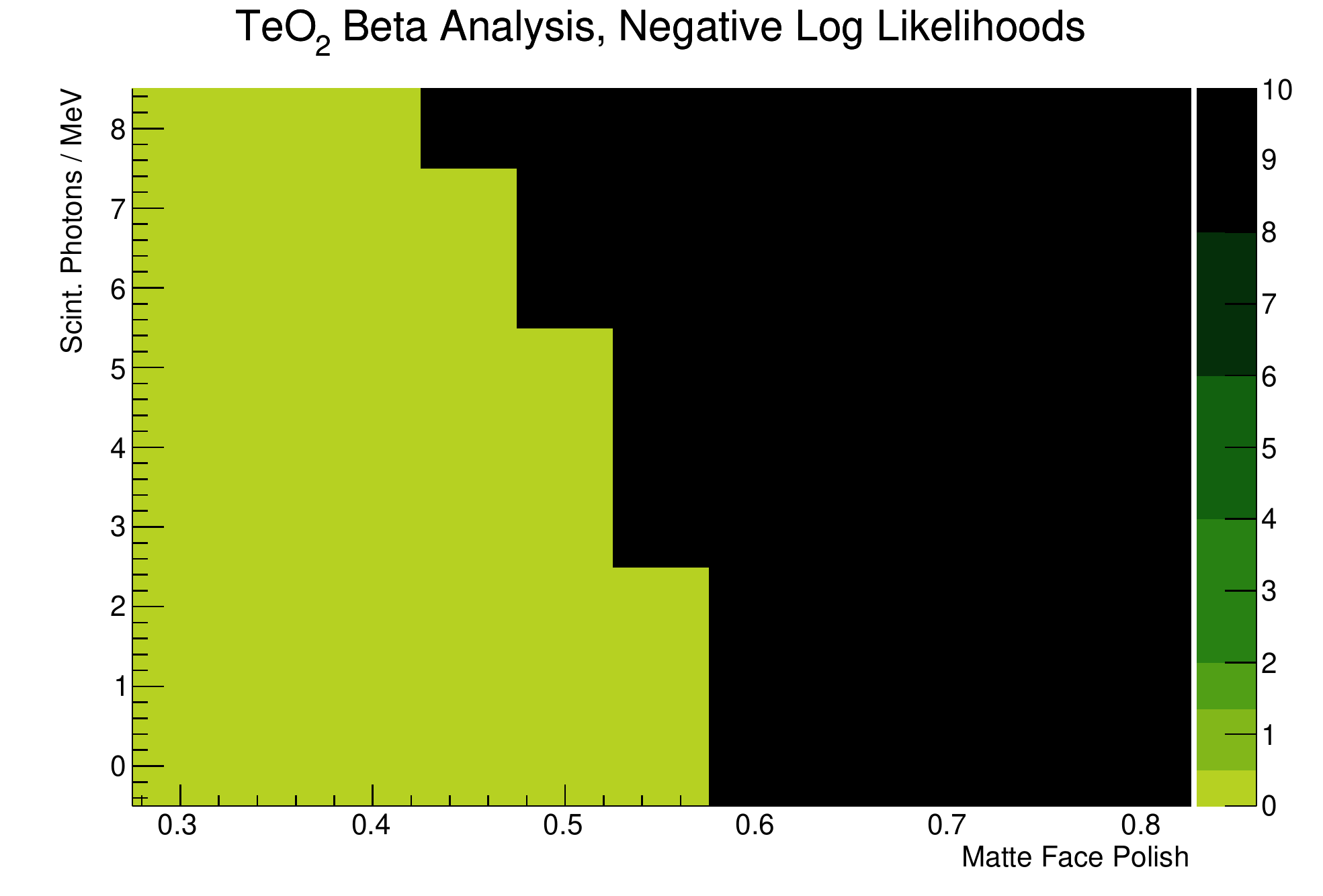}
  \includegraphics[width=0.32\textwidth]{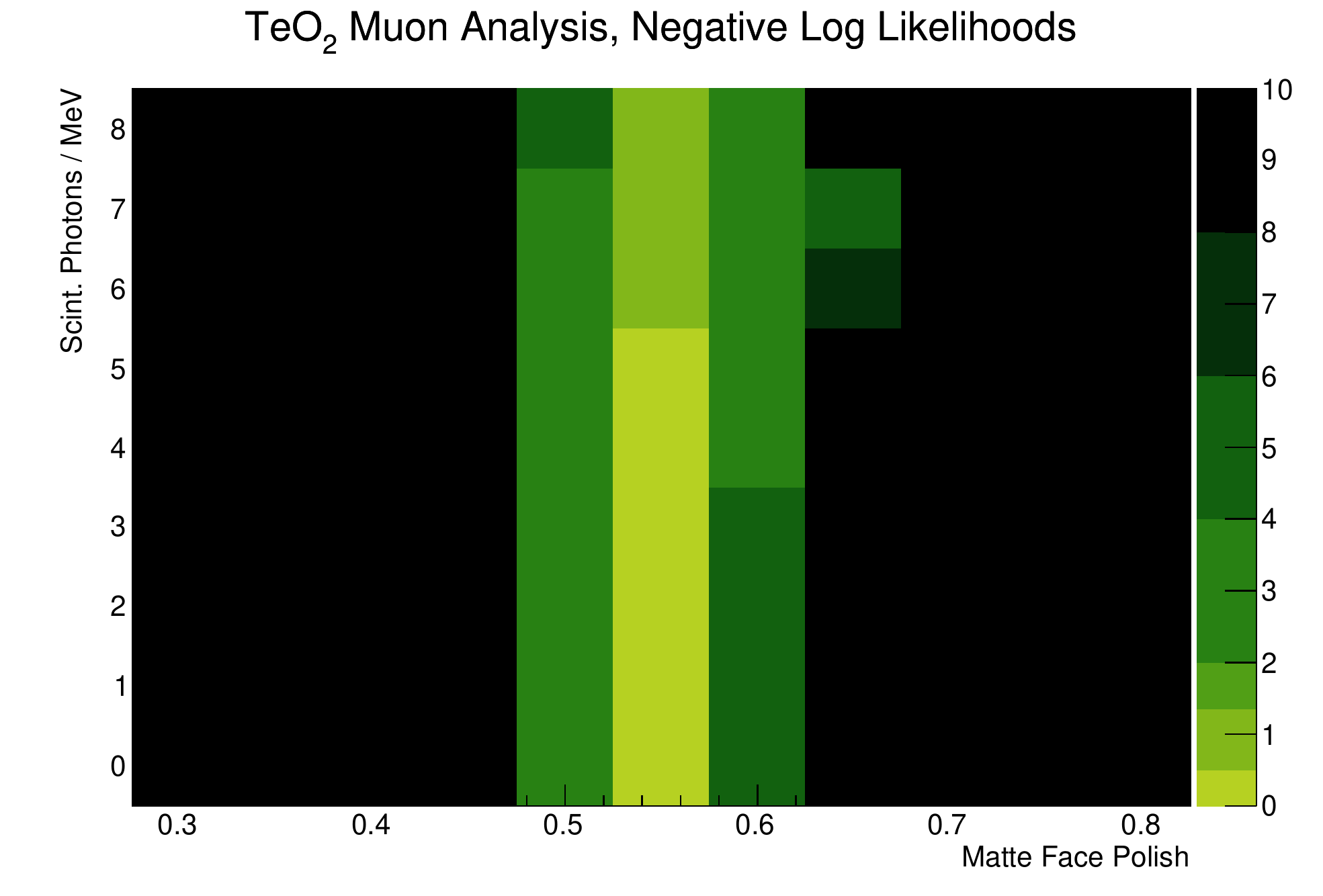}
  \includegraphics[width=0.32\textwidth]{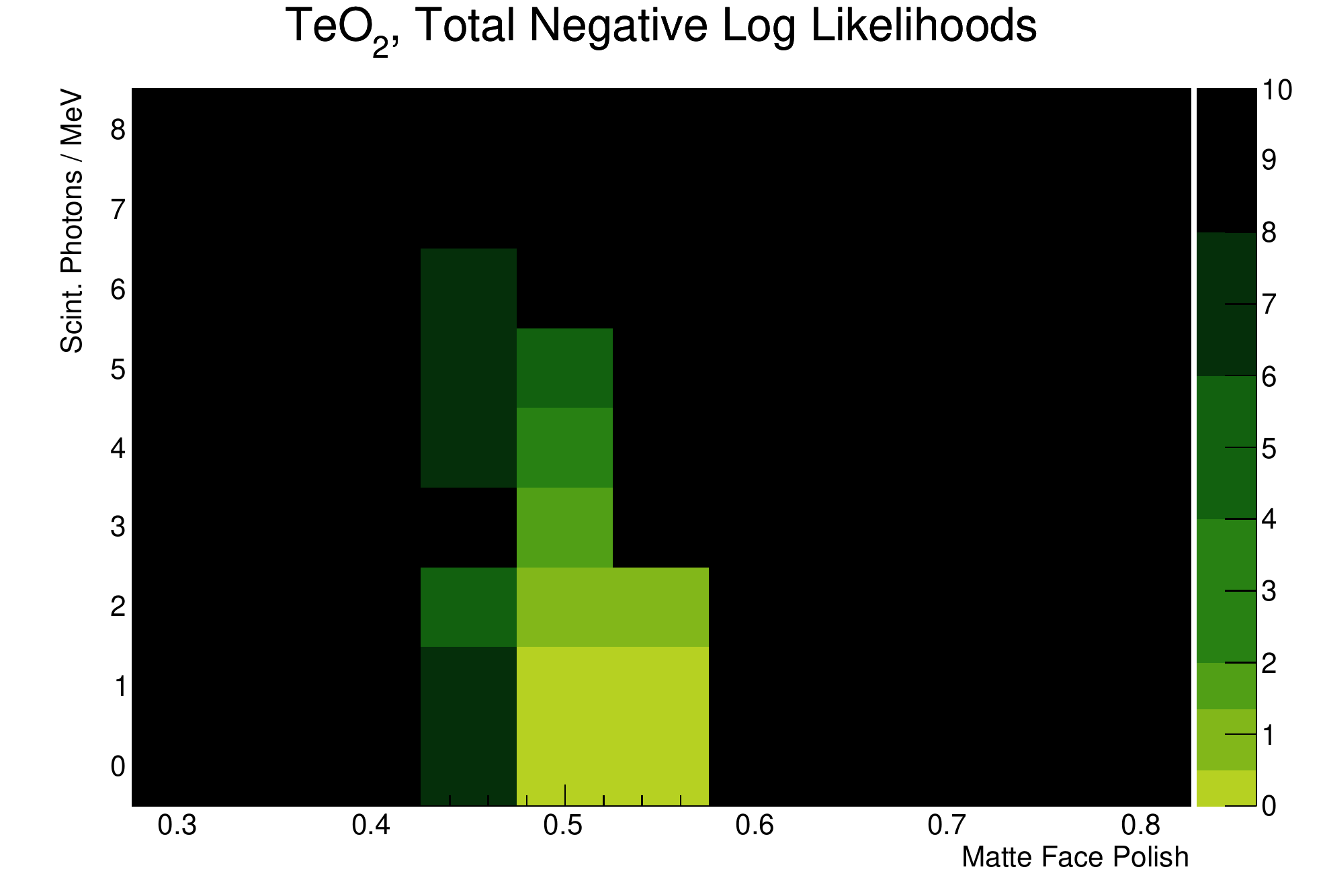}
  \caption{Negative log likelihood for different matte face polish and scintillation yield settings in TeO$_2$, with each box having been profiled over the glossy face polish. Left: $\beta$ analysis results, where it can be seen that there is some degeneracy in the effects of adjusting the surface polish and the scintillation yields. The additional degeneracy compared to the $\beta$ plot in Fig. \ref{fig:UVT_Contour} is due to the presence of a second polish parameter in the \teo analysis. Middle: muon data, which mostly constrains the polish settings, as it did in the UVT acrylic analysis. Right: the total negative log likelihoods obtained by combining the $\beta$ and muon analyses. Each plot is normalized to have a minimum negative log likelihood of 0 and does not include systematic uncertainties.}
  \label{fig:TeO2_Contour}
\end{figure*}

\begin{figure}
  \centering
  \includegraphics[width=0.7\columnwidth]{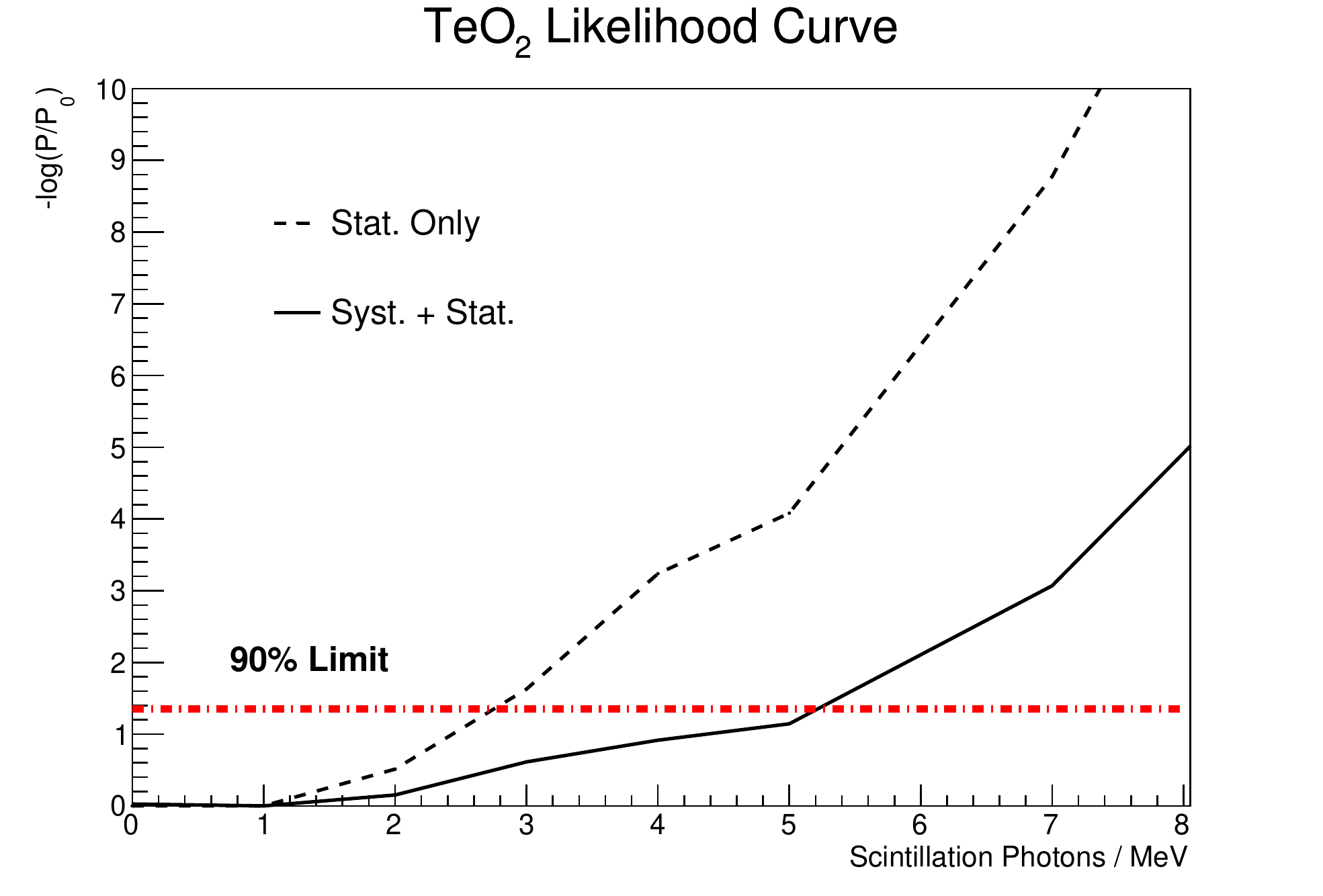}
  \caption{Negative log likelihood curve for the scintillation yield of \teo plotted both with and without systematic uncertainties, normalized to have a best fit of 0 log likelihood. We obtain a 90\% upper limit of 5.3 scintillation photons / MeV in \teo.}
  \label{fig:TeO2_NLL}
\end{figure}

\begin{figure}
  \centering
  \includegraphics[width=0.65\columnwidth,page=1]{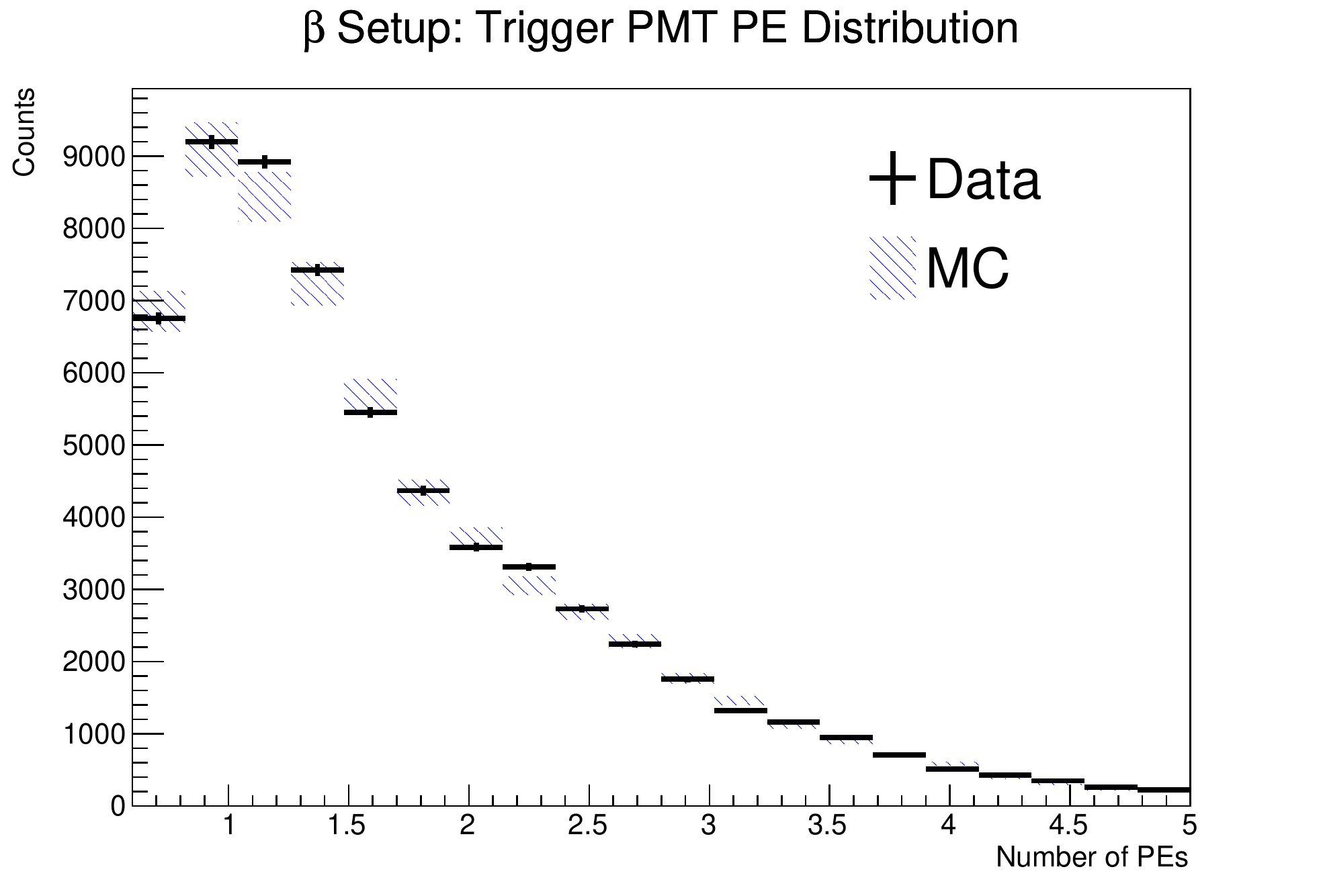}
  \includegraphics[width=0.45\columnwidth,page=2]{figs/TeO2BestFits.pdf}
  \includegraphics[width=0.45\columnwidth,page=3]{figs/TeO2BestFits.pdf}
  \caption{Distributions for the nominal best fit (statistical error only) for the \teo $\mu$/$\beta$ analysis. This minimized negative log likelihood corresponds to a matte face polish of 0.55, a glossy face polish of 0.85, and a scintillation yield of 0 photons / MeV, as shown in Fig. \ref{fig:TeO2_Contour}. Shaded Monte Carlo regions correspond to $1\sigma$ systematic uncertainties. Top: PE distribution in trigger PMT 2, used for $\beta$ event counting. Bottom left: Ratio of number of PEs in the middle ring to the number in the inner ring for cosmic muons. Bottom right: Ratio of number of PEs in the inner ring to the number in the outer ring for cosmic muons.}
  \label{fig:TeO2_BestFits}
\end{figure}

The effects of polish and scintillation yield on the negative log likelihood analysis for \teo can be seen in Fig. \ref{fig:TeO2_Contour}, where the results have been profiled over the glossy face polish and include only statistical uncertainties. In a similar fashion to the acrylic analysis, there is a degeneracy in the effects of surface polish and increased scintillation yields when looking only at the $\beta$ data, but the muon data provides strong constraints on the surface polish settings. The corresponding delta log likelihood curve for \teo is shown in Fig. \ref{fig:TeO2_NLL}, with a 90\% confidence upper limit of 5.3 scintillation photons / MeV. In comparison to the expected 105 to 108 photons / MeV from Cherenkov light \cite{Casali2017}, this corresponds to a less than 5\% contribution to the total light yield from scintillation, setting a stronger limit compared to previous work in \cite{Casali2013}. The data overlaid with the Monte Carlo-predicted distributions for the nominal best fit polish and scintillation yields are shown in Fig. \ref{fig:TeO2_BestFits}.

\subsection{Alpha scintillation yield}

\begin{figure}
  \centering
  \includegraphics[width=0.7\columnwidth]{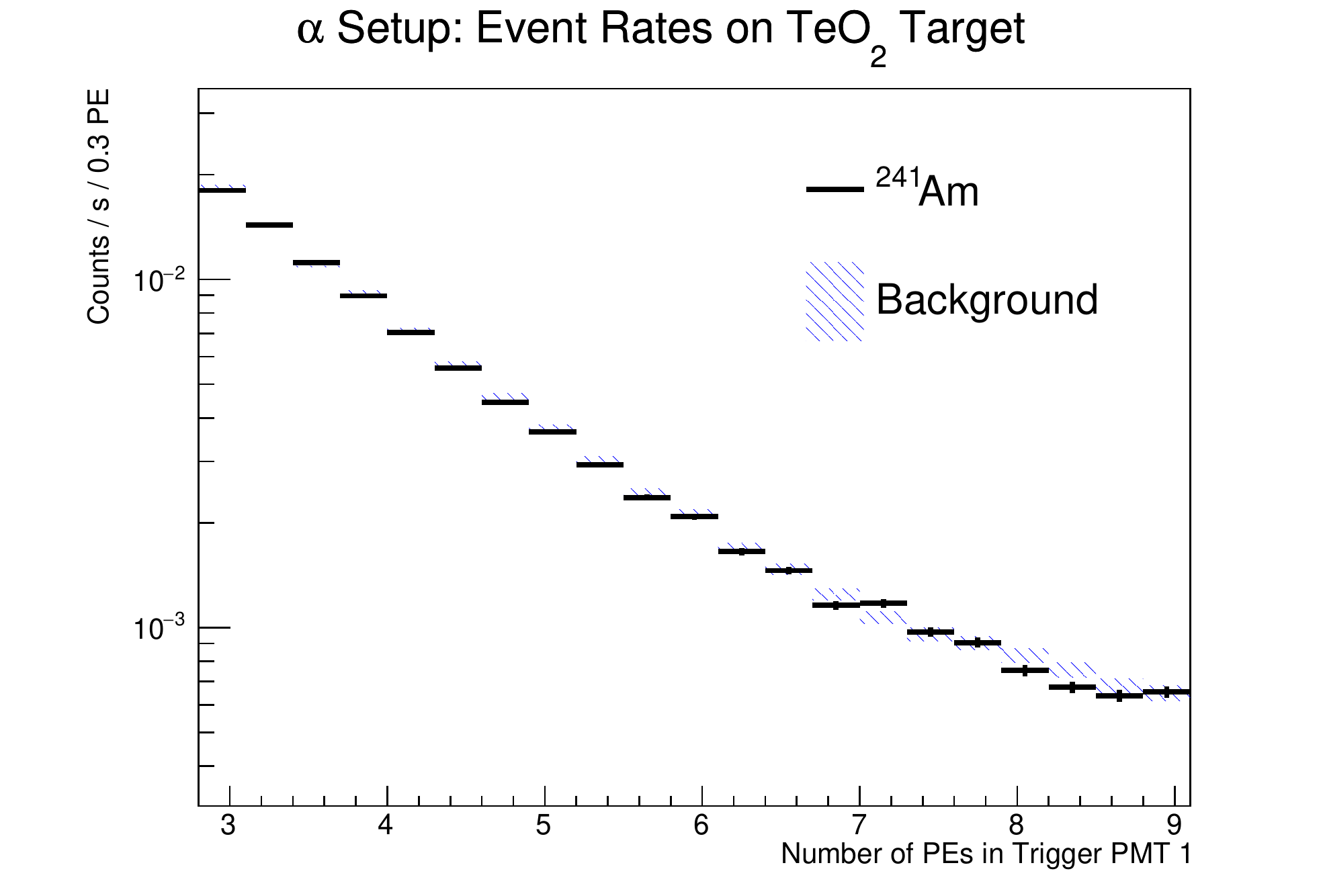}
  \caption{Measured PE distribution in trigger PMT 1 after all cuts are applied in the radioactive-source configuration, for both the presence of an $\alpha$ source and the presence of no source.}
  \label{fig:TeO2_AlphaBgDistribution}
\end{figure}

\begin{figure}
  \centering
  \includegraphics[width=0.7\columnwidth]{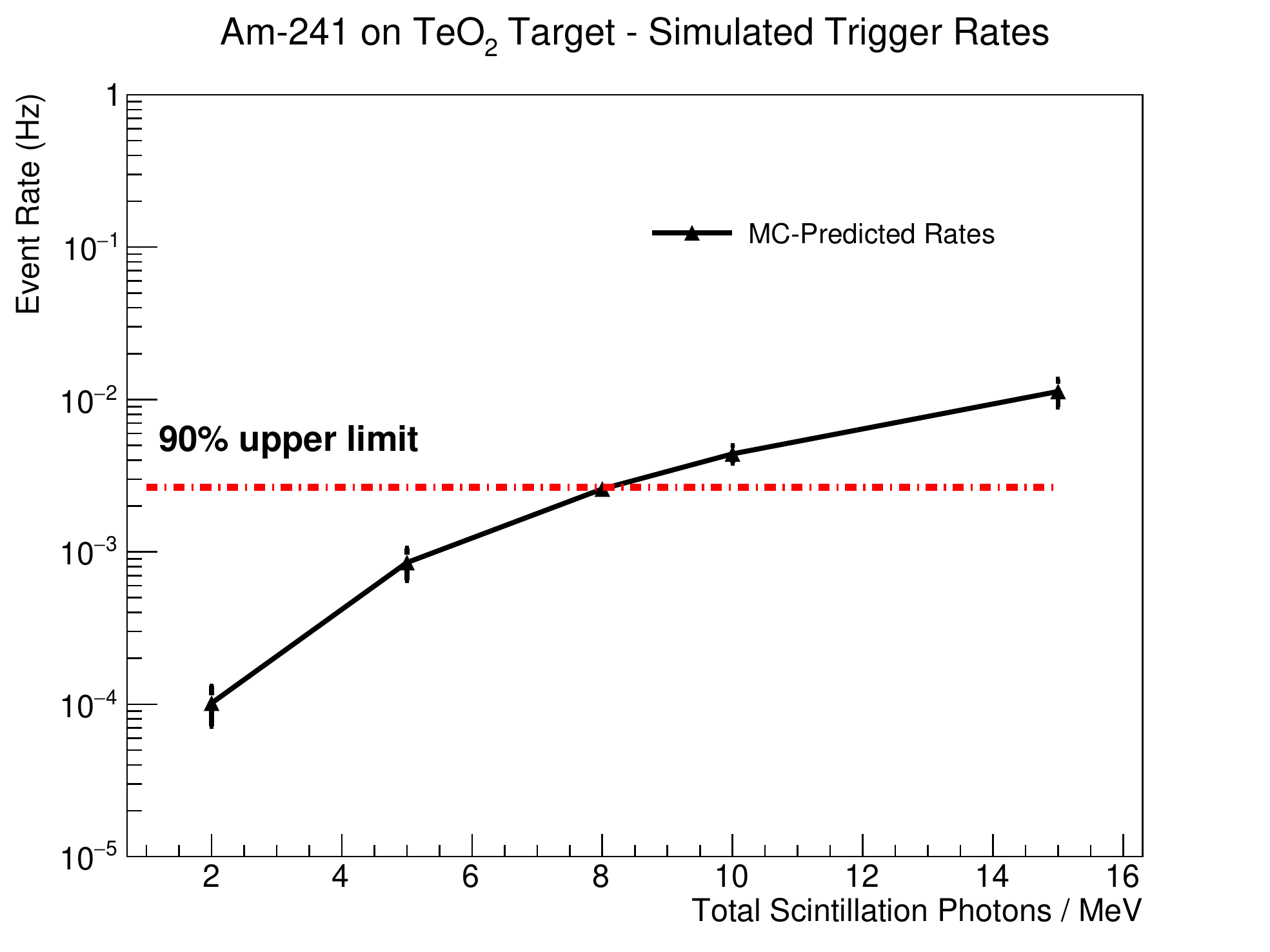}
  \caption{Monte Carlo predictions for the event rate in the $\alpha$ light-yield setup as a function of the $\alpha$ scintillation yield. The 90\% upper limit on the measured $\alpha$ rate is drawn as a horizontal line, corresponding to a limit of $<8$ scintillation photons / MeV.}
  \label{fig:TeO2_AlphaLimit}
\end{figure}

We find that the $\alpha$ and background data are statistically compatible, as seen in the PE distribution in trigger PMT 1 shown in Fig. \ref{fig:TeO2_AlphaBgDistribution}. We measure an event rate of $0.1610 \pm 0.0004 \text{ (stat.)} \pm 0.0006 \text{ (sys.)}$ Hz with the $\alpha$ source deployed and an event rate of $0.1609 \pm 0.0005 \text{ (stat.)} \pm 0.0014 \text{ (sys.)}$ Hz as background after all offline cuts are applied, for an overall background-subtracted $\alpha$ event rate of
\[(0.1 \pm 0.6\text{ (stat.)} \pm 1.5\text{ (sys.)}) \times 10^{-3} \text{ Hz}\]
This gives a 90\% confidence upper limit on the $\alpha$ event rate of $2.8 \times 10^{-3}$ Hz, corresponding to a limit on $\alpha$ scintillation in \teo of 8 photons / MeV as shown in Fig. \ref{fig:TeO2_AlphaLimit}. This limit confirms that any scintillation light from $\alpha$s in \teo is negligibly small at room temperature, though there remains the possibility of some small contribution that appears at cryogenic temperatures.

